\documentclass[10pt,prl,twocolumn,natbib,showkeys] {revtex4}

\usepackage[dvips]{graphicx}

\begin{document}

\title{Influence of the Josephson junction geometry on the spectral properties of the flux-flow oscillator}

\author{Leonid S. Revin$^{1,2}$,
Andrey L. Pankratov$^{1,2,}$}
\email{alp@ipm.sci-nnov.ru}

\affiliation{$^1$Institute for Physics of Microstructures of RAS,
GSP-105, Nizhny Novgorod, 603950, Russia
\\$^2$Laboratory of Cryogenic Nanoelectronics, Nizhny Novgorod State Technical University, Nizhny Novgorod, Russia}

\begin{abstract}
Numerical investigation of the effects of junction geometry on the performance of the flux-flow type Josephson oscillator has been done. Using direct simulation of the sine-Gordon equation with account of RC load the spectral and power properties for different current density distributions have been calculated. With discussion of the fundamental differences in the dynamics of the inline and overlap junctions the various behavior depending on the junction length and noise intensity has been shown.
\end{abstract}
\date{\today}
\maketitle

Flux-flow oscillator (FFO) \cite{begin}, based on a long linear Josephson junction, is presently considered as the most promising local oscillator for superconducting spectrometers \cite{Kosh}. For this purpose the spectral linewidth can be reduced to 1-10 MHz in the 450-700 GHz frequency range. On the other hand, a noisy non-stationary spectrometer \cite{vks} can be created on the basis of the FFO as well. In this case, presently existing FFO designs allow reaching a linewidth of more than 50 MHz at generation frequency from 450 to 700 GHz.

The dynamical and spectral properties of the FFO were investigated both experimentally \cite{begin},\cite{Kosh},\cite{SP}-\cite{KS05} and theoretically \cite{Fiske},\cite{RC}-\cite{MJ}. However, experimental and theoretical investigations aimed to improving spectral and power characteristics were focused on the overlap geometry of Josephson tunnel junctions (JTJ) only. Although in Refs. \cite{PK},\cite{linewidth} the nonuniform current distributions close to the real experimentally observed profiles were studied, orientation of the bias current to the long direction of the junction remained intact. The only analysis of inline character and comparison for different geometries were devoted to the studies of a single fluxon motion \cite{flux1}-\cite{lock}, escape from the zero-voltage state \cite{escape}, Fiske modes \cite{Fiske} and the current-voltage (IV) dependence of the FFO regime \cite{FFOin},\cite{FFOinov}. The oscillation spectra and the linewidth of the FFO in the inline geometry has not been studied ether experimentally or theoretically.

The main goal of the present paper is to study the influence of the Josephson junction geometry on the flux-flow regime and to compare spectral and power properties of the inline and overlap bias feeds.

It is known, that all basic properties of the FFO including spectral characteristics can adequately be described in the frame of the sine-Gordon equation:
\begin{equation}
\phi_{tt} + \alpha\phi_t - \phi_{xx} = \beta\phi_{xxt} + \eta(x) - \sin(\phi) + \eta_f(x,t) \label{sine-Gordon}
\end{equation}
where indices $t$ and $x$ denote temporal and spatial derivatives, $\phi$ is the phase order parameter. Space and time are normalized to the Josephson penetration length $\lambda_J$ and to the inverse plasma frequency $\omega_p^{-1}$, respectively, $\alpha = \omega_p/\omega_c$ is the damping parameter, $\omega_p = \sqrt{2eI_c/\hbar C}$, $\omega_c = 2eI_cR_N/\hbar$, $I_c$ is the critical current, $C$ is the junction capacitance, $R_N$ is the normal state resistance, $\beta$ is the surface loss parameter, $\eta(x)$ is the injected current density, normalized to the critical current density $J_c$, and $\eta_f(x,t)$ is the fluctuational current density. If the critical current density is fixed and the fluctuations are treated
as white Gaussian noise with zero mean, its correlation function is: $\left<\eta_f(x,t)\eta_f(x',t')\right> = 2\alpha\gamma\delta(x-x')\delta(t-t')$, where $\gamma = I_T/(J_c\lambda_J)$ is the dimensionless noise intensity, $I_T=2ekT/\hbar$ is the thermal current, $e$ is the electron charge, $\hbar$ is the Planck constant, $k$ is the Boltzmann constant and $T$ is the temperature.

It is well known, that the mode of the current injection into the Josephson junctions can affect their static and dynamic properties \cite{lik}. For the overlap geometry, where the bias current is injected into the junction perpendicular to its long direction, most theoretical models have assumed uniform distribution of the bias current along the junction length. However, in a long narrow JTJ the current density distribution across the width of a superconducting electrode film is essentially nonuniform. This distribution has singularities at the junction edges \cite{lik},\cite{types}. In the inline case, where the bias current is injected into the junction parallel to its long direction (which in one-dimensional model leads to injection at the ends of the junction only), the dynamic picture is quite different. Various junction geometries provide the following current density distributions $\eta(x)$ \cite{types}:
\begin{eqnarray}
&\eta_{un}(x)=\eta_0, &\qquad (uniform) \label{uniform} \nonumber\\
&\eta_{ov}(x)=(\eta_0L/\pi)/\sqrt{x(L-x)}, &\qquad (mixed) \label{mixed} \\
&\eta_{in}(x)=\eta_0L[\delta(x)+\delta(x-L)], &\qquad (inline) \label{inline} \nonumber\end{eqnarray}
where $\eta_0$ is a constant given by the total component of the current in the film. The boundary conditions, that simulate simple RC-loads, see Refs. \cite{RC},\cite{PK}-\cite{SelfPump}, have the form:
\begin{eqnarray}
\phi_{x}(0,t)+r_Lc_L\phi_{xt}(0,t)-c_L\phi_{tt}(0,t)+ \nonumber\\
\beta r_Lc_L\phi_{xtt}(0,t) + \beta \phi_{xt}(0,t)=\Gamma,  \label{Left} \\
\phi_{x}(L,t)+r_Rc_R\phi_{xt}(L,t)+c_R\phi_{tt}(L,t)+ \nonumber\\
\beta r_Rc_R\phi_{xtt}(L,t)+\beta \phi_{xt}(L,t)=\Gamma,  \label{Right}
\end{eqnarray}
here $\Gamma$ is the normalized magnetic field, and $L$ is the dimensionless length of the FFO in units of $\lambda_J$. The dimensionless capacitances and resistances, $c_{L,R}$ and $r_{L,R}$, are the FFO RC-load placed at the left and at the right ends, respectively.

Due to the fact that in the inline case the bias current contributes at the ends of the junction only, the inline current density distribution of Eq. (\ref{mixed}) can be accounted for in the boundary conditions instead of the sine-Gordon equation (\ref{sine-Gordon}), where the term $\eta(x)$ is absent (for brevity here we do not write the surface losses and RC-load system as it is done in Eqs. (\ref{Left}),(\ref{Right})): $\phi_x(0,t) =\Gamma-\eta_0L/2$, $\phi_x(L,t)=\Gamma+\eta_0L/2$.

At high magnetic fields a dense chain of vortices under the influence of a bias current is moving unidirectionally through the junction. This fluxon train continuously penetrate from one edge of the junction and radiate at the other. In the overlap JTJ permanent influence of the Lorentz force is maintained through the uniformly distributed bias current, while in the inline geometry fluxons receive an accelerating input due to $\eta_{in}$ near the two junction ends only. Away from the junction edge this inline current influence decreases and for larger damping coefficients this effect is reduced more rapidly. As it was done for a single fluxon motion \cite{flux2},\cite{types}, we can crudely estimate the dependence of the FFO regime established for a different junction parameters using the following terms: in the case $\alpha L$$\ll$$1$ the FFO regime in the inline JTJ is similar to the overlap case. If $\alpha L$$\ge$$1$ the current-voltage dependence as well as spectral properties become different for various bias feeds, Eq. (\ref{mixed}).

\begin{figure}[h]
\resizebox{1\columnwidth}{!}{
\includegraphics{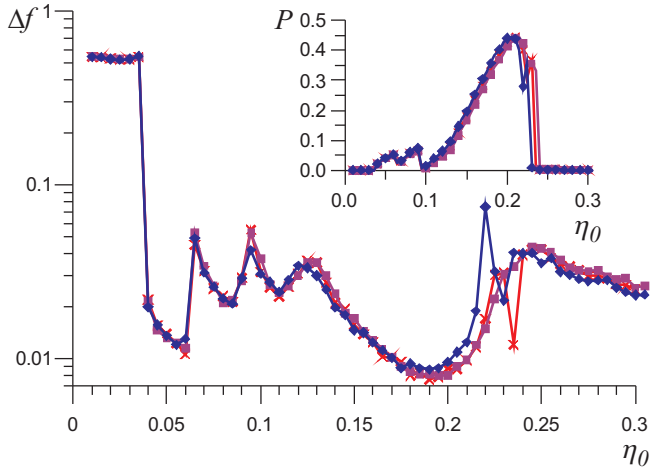}}
{\caption{\small{FFO linewidth versus total bias current, computed for $L=5$ and different $\eta(x)$, see Eq. (\ref{mixed}): curve with rectangles corresponds to uniform distribution of overlap geometry $\eta_{un}$, curve with crosses - mixed overlap $\eta_{ov}$ and curve with diamonds - inline $\eta_{in}$. Inset: Radiated power from the end $x$ = 0 vs $\eta_0$, the notations are the same as for the linewidth.}} \label{Spectr_short}}
\end{figure}

The computer simulations are performed at the same parameters, as in work \cite{linewidth}: $\alpha$=0.033, $\beta$=0.035, $\Gamma$=3.6, $c_L=c_R$=100, $r_R$=2, $r_R$=100, but for lower noise intensity $\gamma$=0.05. Firstly let us consider a rather short JTJ with $L$ = 5. The IV curves we obtained were similar to the results of Ref. \cite{SelfPump}. The FFO linewidth for different bias current profiles is presented in Fig. \ref{Spectr_short}, while the power at RC load - at the inset of Fig. \ref{Spectr_short}. As discussed above, for inline junctions $\alpha L<1$ the vortices starting to move from the input edge under the boundary influence of a bias current, do not have a large speed reduction in comparison with overlap JTJ, so FFO regimes established are the same with similar spectral and power characteristics.

\begin{figure}[h]
\resizebox{1\columnwidth}{!}{
\includegraphics{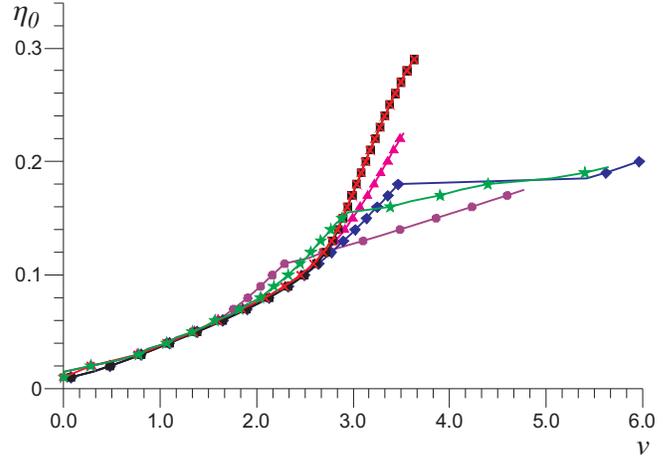}}
{\caption{\small{The IVC of loaded FFO. Uniform overlap (curves coincide): $L=20$ (curve with crosses) and $L=40$ (curve with rectangles), mixed overlap: $L=20$ (curve with triangles) and $L=40$ (curve with diamonds), inline: $L=20$ (curve with stars) and $L=40$ (curve with circles). }} \label{IV_long}}
\end{figure}
Now let us consider more interesting from a practical point of view case of a long junction $\alpha L>1$. The IV characteristics for different junction geometries and $L=20$, $L=40$ are presented in Fig. \ref{IV_long}. As it has been shown in Ref. \cite{FFheight}, for uniform current distribution the dependence of the flux-flow step height $\eta_{0h}$ on the normalized junction length $L$ is nonlinear: for small lengths the step height $\eta_{0h}$ increases with $L$ and then becomes almost constant. As one can see from Fig. \ref{IV_long}, the IV curves for current density $\eta_{un}$ actually coincide. While for mixed overlap and inline cases the step heights for $L=20$ are larger than for $L=40$, i.e. from a certain $L$ values increase of junction length leads to decrease of step heights. In a very long inline junctions the flux-flow step may disappear entirely. This is because the energy input from such long-distance edges does not allow to create a regime of vortex train flow.

The maximum height of the current step $\eta_{0h}$ corresponds to voltage $v_h$ which can be determined from the velocity-matching condition \cite{FFheight}, so that $v_h \cong \Gamma$. For different distributions $\eta_{un}$, $\eta_{ov}$ and different lengths computer simulations agrees well with the obtained analytical relation, see Fig. \ref{IV_long}. But, in the inline character the flow velocity varies spatially \cite{FFOinov}, and the longer the junction, the greater this change. Therefore, even for different lengths of inline junctions the voltages $v_h$ are different.

\begin{figure}[h]
\resizebox{1\columnwidth}{!}{
\includegraphics{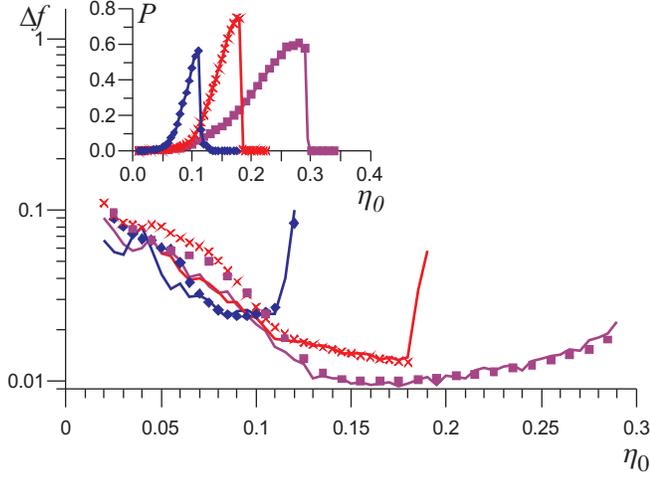}}
{\caption{\small{The FFO linewidth for $L=40$ and different junction geometries: uniform overlap (violet curve and rectangles - simulations and theory (\ref{lwffo})), mixed overlap (red curve and crosses - simulations and theory (\ref{lwffo})), inline (blue curve and diamonds - simulations and theory (\ref{lwffo})). The inset: the radiated power (simulations) for the same geometries.}} \label{Spectr_long}}
\end{figure}
Now let us look at the spectral and power characteristics for junctions with $L=40$. The largest signal in the flux-flow regime is obtained biasing the junction near the top of the flux-flow branch $\eta_{0h}$. Although the step height for various junction geometries is different, the maximum radiated power for all three cases is almost the same and corresponds to their own $\eta_{0h}$, see the inset of Fig. \ref{Spectr_long}. The linewidths of the FFOs at the largest signal are: $\Delta f_{un}=0.0179$, $\Delta f_{ov}=0.0133$, $\Delta f_{in}=0.0247$, while the minimal attainable linewidths are $\Delta f_{un}=0.0094$, $\Delta f_{ov}=0.0133$, $\Delta f_{in}=0.0243$. So, for the inline JTJ the linewidth is 2-2.5 times larger than for the overlap uniform and mixed current distributions. Therefore, the inline junctions seem to be more suitable for an application of a noisy FFO with a broad linewidth and quasi-monochromatic generation. However, special designs of the overlap junction electrodes can possibly lead to larger linewidths than the inline junctions have \cite{KS03}, but this subject is out of scope of the present paper.

In Ref. \cite{formul}, the formula for the FFO linewidth, that takes into account not only conventional differential resistance over bias
current $r_d=dv/d\eta_0$, but also differential resistance over magnetic field $r_d^{CL}=Ldv/d\Gamma$ (control line current) was derived:
\begin{equation}
\Delta f_{FFO}=2\alpha\gamma (r_d+\sigma r^{CL}_d)^2 /L,
\label{lwffo}
\end{equation}
and demonstrated good agreement with experiment \cite{KSsust},\cite{exper}, and also with computer simulation results for overlap junctions \cite{linewidth},\cite{SelfPump}. In Fig. \ref{Spectr_long} the results of computer simulations of the linewidth for different bias feeds are compared with the theory (\ref{lwffo}) with the only fitting parameter $\sigma$ (violet curve and rectangles - simulations and theory (\ref{lwffo}) with $\sigma=0.205$ for uniform overlap bias, red curve and crosses - simulations and theory (\ref{lwffo}) with $\sigma=0.06$ for mixed overlap bias, and blue curve and diamonds - simulations and theory (\ref{lwffo}) with $\sigma=0.01$ for inline bias). One can see that theory (\ref{lwffo}) demonstrates good agreement with the computer simulation results for all three cases, and that the appropriate value of $\sigma$ decreases with increase of the linewidth, so $\sigma$ is minimal for the inline case. Certain disagreement in the area of small and large $\eta_0$ is due to the fact that away from the flux flow step the IV curves for different values of $\Gamma$ actually coincide, and it is impossible to correctly calculate $r_d^{CL}$ in this range.

\begin{figure}[h]
\resizebox{1\columnwidth}{!}{
\includegraphics{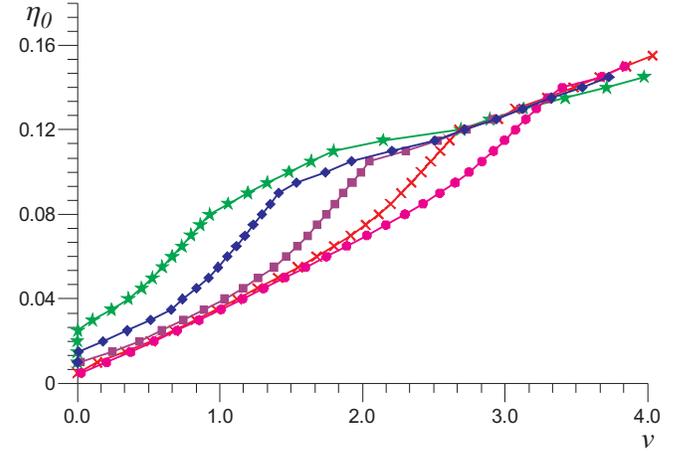}}
{\caption{\small{The IVC of loaded FFO, inline geometry, $L=40$. Curve with stars - $\Gamma=1.5$, curve with diamonds - $\Gamma=2.2$, curve with rectangles - $\Gamma=3.2$, curve with crosses - $\Gamma=4.2$, curve with circles - $\Gamma=5.2$. }} \label{IV(G)}}
\end{figure}
\begin{figure}[h]
\resizebox{1\columnwidth}{!}{
\includegraphics{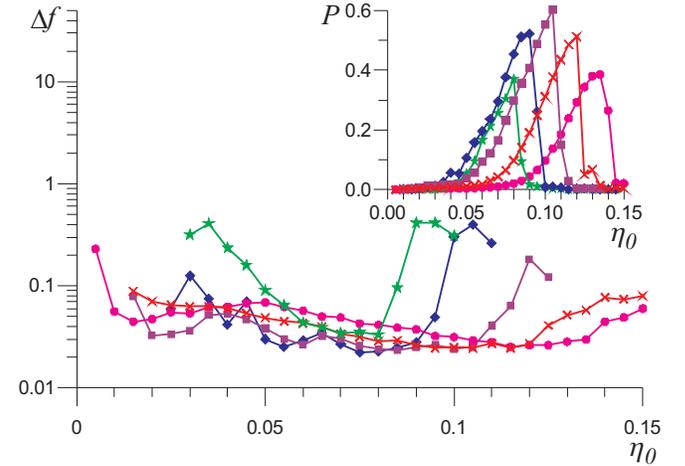}}
{\caption{\small{Linewidth and radiated power (the inset) of the inline FFO, $L=40$. Curve with stars - $\Gamma=1.5$, curve with diamonds - $\Gamma=2.2$, curve with rectangles - $\Gamma=3.2$, curve with crosses - $\Gamma=4.2$, curve with circles - $\Gamma=5.2$. }} \label{Spectr(G)}}
\end{figure}
It is well known \cite{begin},\cite{Kosh},\cite{SP},\cite{PK} that in the overlap JTJ reducing of the magnetic field ($\Gamma \le 2.5$) leads to transformation of the flux-flow regime into Fiske steps, where is no possibility of continuous frequency tuning. In Fig. \ref{IV(G)} the IV curves for inline junction $L=40$ depending on the external magnetic field $\Gamma$ are presented. The FFO regime is stored in the broad range $\Gamma \ge 1.5$ (the Fiske steps are almost invisible even in the limit of vanishing noise intensity and continuous frequency tuning seems to be possible) and the minimal attainable linewidth for each $\Gamma$ in the working areas is roughly the same, see Fig. \ref{Spectr(G)}. The maximum radiated power slightly varies versus $\Gamma$ and the largest signal corresponds to $\Gamma \approx 3.2$.

\begin{figure}[h]
\resizebox{1\columnwidth}{!}{
\includegraphics{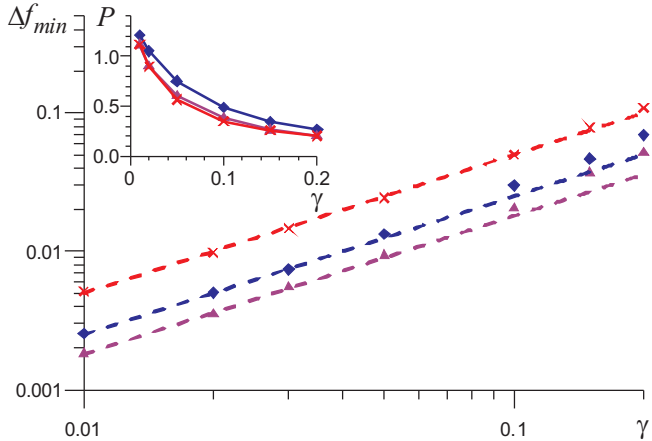}}
{\caption{\small{Minimal attainable linewidth and maximal radiated power (the inset) for $L=40$, $\Gamma=3.6$ and different junction geometries: uniform overlap (curve with triangles), mixed overlap (curve with diamonds), inline (curve with crosses). Dashed lines - linear fit to computed data.}} \label{fP(gamma)}}
\end{figure}
The influence of noise on the FFO spectral properties is estimated by plotting the minimal attainable linewidth and the maximal power values for different current density distributions, see Fig. \ref{fP(gamma)}. It is seen that the effect of noise on inline JTJ in comparison with overlap character results both in the larger linewidth and in the steeper slope of the $\Delta f_{min}(\gamma)$ curve. For the uniform and mixed overlap cases the slopes are $0.18\gamma$ and $0.25\gamma$, respectively, while for inline junction it corresponds to $0.5\gamma$ dependence. Also, one can see the deviation from linear dependence of the linewidth starting from $\gamma\ge 0.1$. Behavior of the radiated power is strictly non-linear: for all geometries it sharply decreases by a factor of three for $\gamma=0.1$ compared with $\gamma=0.01$.

In conclusions, the spectral and power properties of a long Josephson tunnel junction have been studied for various junction geometries by numerical solution of the one-dimensional sine-Gordon equation taking into account surface losses and RC load. The fundamental differences in the dynamics of the FFO with various current density distributions have been discussed and the distinct behavior of inline and overlap junctions with changing the external magnetic field and noise intensity is shown. Lower performance of the inline junction (compared to the overlap) as a local oscillator is observed. On the other hand the broad range of magnetic fields, where continuous frequency tuning is possible, and a large linewidth allows one to consider inline junctions as an application for a noisy non-stationary FFO spectrometer.

 The work was supported by RFBR Project No. 09-02-00491, Human Capital Foundation, Dynasty Foundation and by the Act 220 of Russian Government (project 25).


\begin{thebibliography}{99}

\bibitem{begin} T. Nagatsuma, K. Enpuku, F. Irie, and K. Yoshida,  J. Appl. Phys. {\bf 54}, 3302 (1983);
{\bf 56}, 3284 (1984); {\bf 58}, 441 (1985); {\bf 63}, 1130 (1988).

\bibitem{Kosh} V.P. Koshelets and S.V. Shitov, Supercond. Sci. Technol. {\bf 13}, R53 (2000).

\bibitem{vks} V.L. Vaks, V.V. Khodos, E.V. Spivak, Rev. Sci. Inst.,  \textbf{70}, 3447 (1999).

\bibitem{SP} V.P. Koshelets, S.V. Shitov, A.V. Shchukin, L.V. Filippenko, J.
Mygind, and A.V. Ustinov, Phys. Rev. B \textbf{56}, 5572 (1997).

\bibitem{Fiske} M. Cirillo, N. Gr{\o }nbech-Jensen, M.R. Samuelsen, M. Salerno, G. Verona Rinati, Phys. Rev. B {\bf 58} 12377 (1998).

\bibitem{KSsust} V. P. Koshelets, P.N. Dmitriev, A.B. Ermakov, A.S. Sobolev, A.M. Baryshev, P.R. Wesselius and J. Mygind,
Supercond. Sci. Technol. {\bf 14} 1040 (2001).

\bibitem{exper} V.P. Koshelets, P.N. Dmitriev, A.S. Sobolev, A.L. Pankratov, V.V. Khodos, V.L. Vaks, A.M. Baryshev, P.R. Wesselius, and J. Mygind, Physica C {\bf 372-376}, 316 (2002).

\bibitem{KS03} V.P. Koshelets, S.V. Shitov, L.V. Filippenko, P.N. Dmitriev,
A.B. Ermakov, A.S. Sobolev, M.Yu. Torgashin, A.L. Pankratov, V.V.
Kurin, P. Yagoubov, and R. Hoogeveen, Supercond. Sci. Technol. {\bf
17}, S127 (2004).

\bibitem{KS05} V.P. Koshelets, P.N. Dmitriev, A.B. Ermakov, A.S. Sobolev, M.Yu. Torgashin,
V.V. Kurin, A.L. Pankratov, and J. Mygind, IEEE Trans. Appl.
Supercond. {\bf 15}, 964 (2005).

\bibitem{RC} C. Soriano, G. Costabile, and R.D. Parmentier, Supercond. Sci. Technol., {\bf 9}, 578 (1996).

\bibitem{ss99}  M. Salerno and M. R. Samuelsen, Phys.\ Rev.\ B, {\bf 59}, 14653 (1999).

\bibitem{jaw}  M. Jaworski, Phys. Rev. B, {\bf 60}, 7484 (1999).

\bibitem{formul} A.L. Pankratov, Phys. Rev. B {\bf 65}, 054504 (2002).

\bibitem{PK} A.L. Pankratov, A.S. Sobolev, V.P. Koshelets, and J. Mygind, Phys. Rev. B {\bf 75}, 184516 (2007).

\bibitem{linewidth} A.L. Pankratov, Appl. Phys. Lett. {\bf 92}, 082504 (2008).

\bibitem{SelfPump} A.L. Pankratov, Phys. Rev. B {\bf 78}, 024515 (2008).

\bibitem{MJ1} M. Jaworski, Supercond. Sci. Technol. {\bf 21}, 065016 (2008).

\bibitem{MJ} M. Jaworski, Phys. Rev. B, \textbf{81}, 224517 (2010).

\bibitem{flux1} O.A. Levring, N.F. Pedersen, and M.R. Samuelsen, Appl. Phys. Lett. {\bf 40}, 846 (1982).

\bibitem{flux2} O.A. Levring, N.F. Pedersen, and M.R. Samuelsen, J. Appl. Phys. {\bf 54}, 987 (1982).

\bibitem{lock} N.F. Pedersen, A. Davidson, Phys. Rev. B {\bf 41}, 178 (1990).

\bibitem{escape} M.G. Castellano, G. Torrioli, C. Cosmelli, A. Costantini, F. Chiarello, P. Carelli, G. Rotoli, M. Cirillo, and R.L. Kautz, Phys. Rev. B {\bf 54}, 15417 (1996).

\bibitem{FFOin} P. Cikmacs, M. Cirillo, V. Merlo, R. Russo, IEEE Trans. Appl. Supercond. {\bf 11(1)} 99 (2001).

\bibitem{FFOinov} K. Yoshida, T. Nagatsuma, K. Sueoka, K. Enpuku and F. Irie, IEEE Trans. Magn. {\bf 21(2)} 899 (1985).

\bibitem{lik}  K. K. Likharev, {\it Dynamics of Josephson Junctions and
Circuits} (Gordon and Breach, New York, 1986).

\bibitem{types} M.R. Samuelsen and S.A. Vasenko, J. Appl. Phys. {\bf 57}, 110 (1985).

\bibitem{flux3} O.H. Olsen, M.R. Samuelsen, J. Appl. Phys. {\bf 54}, 2777 (1982).

\bibitem{FFheight} Y.M. Zhang, P.H. Wu, J. Appl. Phys. {\bf 68}, 4703 (1990).

\end{thebibliography}
\end{document}